\documentclass[showpacs, showkeys,twocolumn]{revtex4}
\begin{document}
\title{ Quark-Antiquark and Diquark Condensates in Vacuum
in a 2D Two-Flavor Gross-Neveu Model\footnote{The project
supported by the National Natural Science Foundation of China
under Grant No.10475113.}}
\author{Zhou Bang-Rong }
\affiliation{College of Physical Sciences, Graduate School of the
Chinese Academy of Sciences, Beijing 100049, China}\affiliation{
CCAST (World Laboratory), P.O.Box 8730, Beijing 100080, China}
\date{}
\begin{abstract}
The analysis based on the renormalized effective potential
indicates that, similar to in the 4D two-flavor Nambu-Jona-Lasinio
(NJL) model, in a 2D two-flavor Gross-Neveu model, the interplay
between the quark-antiquark and the diquark condensates in vacuum
also depends on  $G_S/H_S$, the ratio of the coupling constants in
scalar quark-antiquark and scalar diquark channel. Only the pure
quark-antiquark condensates exist if $G_S/H_S>2/3$ which is just
the ratio of the color numbers of the quarks participating in the
diquark and quark-antiquark condensates.  The two condensates will
coexist if $0<G_S/H_S<2/3$. However, different from the 4D NJL
model, the pure diquark condensates arise only at $G_S/H_S=0$ and
are not in a possibly finite region of $G_S/H_S$ below 2/3.
\end{abstract}
\maketitle
\section{Introduction\label{Intro}}
It is well known that the four-fermion interaction models
including the Nambu-Jona-Lasinio (NJL) model \cite{kn:1} in 4D
space-time and the Gross-Neveu (GN) models \cite{kn:2} in 2D and
3D space-time are useful means to explore dynamical symmetry
breaking. The NJL model may be used in research not only on chiral
symmetry breaking in Quantum Chromodynamics (QCD) which is
connected to the quark-antiquark condensates
\cite{kn:1,kn:3,kn:4,kn:5,kn:6} but also on color superconducting
at low temperature and moderate baryonic density which is related
to the diquark condensates \cite{kn:7}. In fact, owing to that the
four-fermion interactions of $(\bar{q}q)^2$-form and $(qq)^2$-form
may be related to each other via the Fierz transformation, in any
four-fermion model, the above two forms of interactions are always
allowed to exist simultaneously. As a result, one will face to
interplay between the quark-antiquark condensates and the diquark
condensates even at temperature $T=0$ and the quark chemical
potential $\mu=0$, i.e. in vacuum. To clarify such interplay
theoretically is certainly interesting for deeper understanding of
feature of the four-fermion interaction
models.\\
\indent In a preceding paper \cite{kn:8}, we have researched a 4D
two-flavor NJL model and proven that the mutual competition
between the quark-antiquark and the diquark condensates does occur
in vacuum and, depending on the ratio of the coupling constants
$G_S$ and $H_S$ in scalar quark-antiquark and scalar diquark
channel, the diquark condensates could either exist alone, or
coexist with the quark-antiquark condensates, or not be formed
completely. Then a natural question is that whether similar
conclusions can be reached from the similar four-fermion
interaction models in other space-time dimensions?  i.e. whether
these conclusions are of generality for this kind of models?  To
answer this question, we will generalize the discussions in
Ref.\cite{kn:8} to a 2D GN model in this paper. For convenience of
making a comparison with the results in the 4D NJL model given in
Ref.\cite{kn:8}, we will take the fermions (still called quarks)
with zero bare-masses, two flavors and three colors and assume
that the diquark condensates will come from the attractive
interactions in the color anti-triplet channel.\\
\indent The paper is arranged as follows. In Sect.\ref{model} we
will give the 2D two-flavor GN model with diquark interactions,
analyze its discrete symmetries and possible spontaneous breaking
induced by the condensates. In Sect.\ref{effptl} we will derive
the renormalized effective potential in the space-time dimension
regularization approach. In Sect.\ref{groundstates}, by means of
the effective potential, the ground states of the model will be
determined and the mutual competition between the quark-antiquark
and the diquark condensates will be expounded. Finally in
Sect.\ref{concl} our conclusions follow.
\section{2D two-flavor Gross-Neveu model with diquark interactions\label{model}}
The Lagrangian of the model is expressed by
\begin{eqnarray}
{\cal L}&=&\bar{q}i\gamma^{\mu}\partial_{\mu} q
+G_S[(\bar{q}q)^2+(\bar{q}i\gamma_5\vec{\tau}q)^2]\nonumber\\
&&+H_S(\bar{q}i\gamma_5\tau_S\lambda_Aq^C)
    ({\bar{q}}^Ci\gamma_5\tau_S\lambda_Aq),
\end{eqnarray}
where, in 2D space-time, $q\equiv q_{\alpha k}(t,x)$ will be two
component spinor fields with the two flavors $\alpha=u, d$ and
three colors $k=r$ (red), $g$ (green), $b$ (blue) and the matrices
$\gamma^{\mu}(\mu=0,1)$, $\gamma_5$ and $C$ will be $2\times2$
matrices which are defined by
\begin{equation}
\gamma^0=\left(
\begin{array}{cc}
  1 & 0 \\
  0 & -1 \\
\end{array}
\right),\;\;\gamma^1=\left(
\begin{array}{cc}
  0 & 1 \\
  -1 & 0 \\
\end{array}
\right)=-C, \;\;\gamma_5=\gamma^0\gamma^1
\end{equation}
and obey the relations
\begin{equation}
\{\gamma^{\mu},\gamma^{\nu}\}=2g^{\mu\nu},\;\;\{\gamma_5,
\gamma^{\mu}\}=0, \;\; C{\gamma^{\mu}}^TC=\gamma^{\mu}.
\end{equation}
The matrix $C$ is related to the charge conjugates of the quark
fields $q$ by
\begin{equation}
q^C=C\bar{q}^T, \;\; \bar{q}^C=q^TC.
\end{equation}
The Pauli matrices $\vec{\tau}=(\tau_1, \tau_2, \tau_3)$ act in
the two-flavor space and $\tau_S=(\tau_0\equiv1, \tau_1, \tau_3)$
are the flavor-triplet symmetric matrices. The $\lambda_A
(A=2,5,7)$ are the color-triplet antisymmetric Gell-Mann matrices
acting in the three-color space. It is indicated that the product
matrix $C\gamma_5\tau_S\lambda_A$ is antisymmetric and the term
with $H_S$ in Eq.(1) corresponds to the diquark interactions which
could lead to scalar color-antitriplet diquark condensates.\\
\indent Since a continuous symmetry can never be spontaneously
broken in 2D space-time by the Mermin-Wagner-Coleman theorem
\cite{kn:9}, we are interested  only in the discrete symmetries of
the model and their possible spontaneous breaking. In this
respect, it is easy to check that the Lagrangian (1) has the the
following discrete symmetries: $R:
q(t,x)\stackrel{R}{\rightarrow}-q(t,x)$, parity $\mathcal{P}:
q(t,x)\stackrel{\mathcal{P}}{\rightarrow}\gamma^0q(t,-x)$, time
reversal $\mathcal{T}:
q(t,x)\stackrel{\mathcal{T}}{\rightarrow}\gamma^0q(-t,x)$, charge
conjugate $\mathcal{C}:
q(t,x)\stackrel{\mathcal{C}}{\rightarrow}q^C(t,x)$, discrete
chiral symmetry $\chi_D:
q(t,x)\stackrel{\mathcal{\chi_D}}{\rightarrow}\gamma_5q(t,x)$,
special parity $\mathcal{P}_1:
q(t,x)\stackrel{\mathcal{P}_1}{\rightarrow}\gamma^1q(t,-x)$, the
center $Z_3^c$ of $SU_c(3)$:
$$q(t,x)\stackrel{Z_3^c}{\rightarrow}\left(
\begin{array}{ccc}
  1 & 0 & 0 \\
  0 & e^{i2\pi/3} & 0 \\
  0 & 0 & e^{i4\pi/3} \\
\end{array}
\right)_cq(t,x)
$$
and the center $Z_2^f$ of $SU_f(2)$:
$$
q(t,x)\stackrel{Z_2^f}{\rightarrow}\left(
\begin{array}{cc}
  1 & 0 \\
  0 & -1 \\
\end{array}
\right)_f q(t,x).
$$
The subscripts $c$ and $f$ indicate that the
corresponding matrices act in flavor and color space
respectively.\\
\indent Assume that the four-fermion interactions could lead to
the scalar quark-antiquark condensates
\begin{equation}
\langle\bar{q}q\rangle=\phi
\end{equation}
which will break the discrete symmetries $\chi_D$ and
$\mathcal{P}_1$, and the scalar color-antitriplet diquark
condensates
\begin{equation}
\langle\bar{q}^Ci\gamma_51_f\lambda_2q\rangle=\delta
\end{equation}
which will break, besides $\chi_D$ and $\mathcal{P}_1$, the
symmetry $Z_3^c$ down to $Z_2^c$ corresponding to $r$ and $g$
color degree of freedom. For the condensates $\delta$, we have
made appropriate transformations respectively in flavor and color
space and rotated it into $\tau_S=\tau_0=1$ and $A=2$ direction.
\section{Renormalization and effective potential\label{effptl}}
The coupling constants $G_S$ and $H_S$ in Eq.(1) are
dimensionless, thus the theory is perturbatively renormalizable.
We will conduct the renormalization operation in the dimension
regularization approach \cite{kn:10}. To this end, on the basis of
the standard procedure, we change the dimension of space-time from
2 to
\begin{equation}
D=2-2\varepsilon
\end{equation}
and write the renormalized Lagrangian by
\begin{eqnarray}
  {\cal L}_D&=&\bar{q}i\gamma^{\mu}\partial_{\mu} q
+G_SM^{2-D}Z_G[(\bar{q}q)^2+(\bar{q}i\gamma_5\vec{\tau}q)^2]\nonumber\\
&&+H_SM^{2-D}Z_H({\bar{q}}^Ci\gamma_5\tau_S\lambda_Aq)
                (\bar{q}i\gamma_5\tau_S\lambda_Aq^C),
\end{eqnarray}
where $M$ is a scale parameter with mass dimension and the
$\gamma^{\mu}$ are now $2^{D/2}\times2^{D/2}$ matrices. The $Z_G$
and $Z_H$ are renormalization constants which will be
appropriately selected to cancel the ultraviolet (UV) divergences
in the loop integrations.  Our main interest lies in interplay
between the two condensates $\phi$ and $\delta$, so we will derive
the renormalized effective potential of the model containing the
order parameters from the two condensates. Define the order
parameters
\begin{equation}
\sigma=-2G_SM^{2-D}Z_G\phi, \;\; \Delta=-2H_SM^{2-D}Z_H\delta,
\end{equation}
then in the mean field approximation \cite{kn:11} which is
equivalent to the linearization of the interaction terms in the
existence of the corresponding condensates \cite{kn:12} i.e.
$$(\bar{q}q)^2\simeq 2\phi\bar{q}q-\phi^2$$
$$(\bar{q}^Ci\gamma_5\lambda_2q)(\bar{q}i\gamma_5\lambda_2q^C)\simeq
\delta(\bar{q}i\gamma_5\lambda_2q^C)+(\bar{q}^Ci\gamma_5\lambda_2q)\delta^*-|\delta|^2$$
and in the Nambu-Gorkov basis \cite{kn:13} with the bispinor
fields
\begin{equation}
\Psi=\frac{1}{\sqrt{2}}\left(
\begin{array}{c}
  q \\
  q^C \\
\end{array}
\right),
\;\;\bar{\Psi}=\frac{1}{\sqrt{2}}\left(\bar{q},\bar{q}^C\right),
\end{equation}
we can write
$$
\mathcal{L}_D=\bar{\Psi}(x)S^{-1}(x)\Psi(x)-\frac{\sigma^2}{4G_SM^{2-D}Z_G}
-\frac{|\Delta|^2}{4H_S M^{2-D}Z_H}.
$$
In the momentum space, the inverse propagator $S^{-1}(x)$ for the
quark fields will be changed into
\begin{equation}
S^{-1}(p)=\left(
\begin{array}{cc}
  \not\!{p}-\sigma & -i\gamma_5\lambda_2\Delta \\
  -i\gamma_5\lambda_2\Delta^* & \not\!{p}-\sigma  \\
\end{array}
\right),\;\; \not\!{p}=\gamma^{\mu}p_{\mu}.
\end{equation}
The effective potential in vacuum corresponding to $\mathcal{L}_D$
can be expressed by
\begin{eqnarray}
  V(\sigma,|\Delta|) &=& \frac{\sigma^2}{4G_SM^{2-D}Z_G}+
  \frac{|\Delta|^2}{4H_SM^{2-D}Z_H} \nonumber\\
   && +i\int\frac{d^Dp}{(2\pi)^D}\frac{1}{2}\mathrm{Tr}\ln
   S^{-1}(p)S_{0}(p),
\end{eqnarray}
where
$$S_0(p)=\frac{1}{\not\!{p}+i\varepsilon}\left(
\begin{array}{cc}
1 & 0 \\
  0 & 1 \\
\end{array}
\right)$$ represents the propagator for massless quark fields (up
to a factor $i$) and the Tr is taken over flavor, color,
Nambu-Gorkov and Dirac spin degree of freedom.\\
\indent It is seen from the structure of the matrix $\lambda_2$
that the blue quarks do not participate in the diquark condensates
and irrelevant to the order parameter $\Delta$, thus  we may
conduct similar derivation to the one made in Ref. \cite{kn:8} and
obtain
\begin{widetext}
\begin{equation}
V(\sigma,|\Delta|)=\frac{\sigma^2}{4G_SM^{2-D}Z_G}+
  \frac{|\Delta|^2}{4H_SM^{2-D}Z_H} +
  i2^{D/2-1}\int\frac{d^Dp}{(2\pi)^D}
  \left(4\ln\frac{p^2-\sigma^2-|\Delta|^2+i\varepsilon}{p^2+i\varepsilon}
  +2\ln\frac{p^2-\sigma^2+i\varepsilon}{p^2+i\varepsilon}\right),
\end{equation}
\end{widetext}
where in the third term of the right-handed side, the factor 4
come from (the flavor number)$\times$(the color number) of the red
and green quarks participating in the diquark condensates, the
factor 2 from the flavor number of the blue quarks and the factor
$2^{D/2}$ is due to the representation's dimension of the
$\gamma^{\mu}$ matrix. For the momentum integrations, we may make
the Wick rotation, use the Euclidean variable $\bar{p^0}=ip^0,\;
\bar{p^i}=p^i \;(i=1,\cdots,D-1)$ and define the integration
\begin{equation}
I(a^2)=\int\frac{d^D\bar{p}}{(2\pi)^D}\ln\frac{\bar{p}^2+a^2}{\bar{p}^2}\;
\mathrm{with}\; I(a^2=0)=0,
\end{equation}
then it is easy to calculate
\begin{eqnarray}
 I(a^2)&=&\int_{0}^{a^2}du^2\frac{d I(u^2)}{du^2}=
 \int_{0}^{a^2}du^2\int\frac{d^D\bar{p}}{(2\pi)^D}\frac{1}{\bar{p}^2+u^2}  \nonumber\\
   &=& \frac{\Gamma(1-D/2)}{(4\pi)^{D/2}}\frac{(a^2)^{D/2}}{D/2}
\end{eqnarray}
By means of Eqs. (14), (15) and (7) we can transform Eq.(13) to
\begin{widetext}
\begin{eqnarray}
V(\sigma, |\Delta|)&=& \frac{\sigma^2}{4G_SM^{2-D}}\left\{
Z_G^{-1}-1-\frac{6G_S}{\pi}\frac{1}{\varepsilon}+1-\frac{2G_S}{\pi}
\left[2\left(\ln\frac{2\pi M^2}{\sigma^2+|\Delta|^2}-\gamma\right)
+\ln\frac{2\pi M^2}{\sigma^2}-\gamma+3\right]\right\}\nonumber \\
  &&+\frac{|\Delta|^2}{4H_SM^{2-D}}\left\{
  Z_H^{-1}-1-\frac{4H_S}{\pi}\frac{1}{\varepsilon}+1-\frac{4H_S}{\pi}
\left[\ln\frac{2\pi
M^2}{\sigma^2+|\Delta|^2}-\gamma+1\right]\right\},
\end{eqnarray}
\end{widetext}
where $\gamma=0.5772$ is the Euler constant. To eliminate the UV
divergent term with the pole $1/\varepsilon$ (up to one-loop
order), we use the minimal substraction scheme \cite{kn:10} and
define the renormalization constants $Z_G$ and $Z_H$ by
\begin{equation}
Z_G=1-\frac{6G_S}{\pi}\frac{1}{\varepsilon},
\end{equation}
\begin{equation}
Z_H=1-\frac{4H_S}{\pi}\frac{1}{\varepsilon}.
\end{equation}
In this way, the renormalized effective potential up to one-loop
order (after the limit $D\rightarrow 2$ is taken) becomes
\begin{eqnarray}
  V(\sigma, |\Delta|)&=& \frac{\sigma^2}{4G_S}
  -\frac{\sigma^2}{2\pi}\left(2\ln\frac{\bar{M}^2}{\sigma^2+|\Delta|^2}
  +\ln\frac{\bar{M}^2}{\sigma^2}+3\right)\nonumber \\
  &&+\frac{|\Delta|^2}{4H_S}
  -\frac{|\Delta|^2}{\pi}\left(\ln\frac{\bar{M}^2}{\sigma^2+|\Delta|^2}+1\right),
\end{eqnarray}
where the denotation
\begin{equation}
\bar{M}^2=2\pi e^{-\gamma}M^2
\end{equation}
has been used.
\section{Ground states\label{groundstates}}
The effective potential $V(\sigma, |\Delta|)$ depends on the two
real order parameters $\sigma$ and $|\Delta|$. Its minimum points
can be found out analytically, then the ground states of the model
will be determined. From the extreme value conditions $\partial
V(\sigma, |\Delta|)/\partial \sigma=0$ and $\partial V(\sigma,
|\Delta|)/\partial |\Delta|=0$ we get the equations that
\begin{equation}
\sigma\left[1-\frac{2G_S}{\pi}\left(2\ln\frac{\bar{M}^2}{\sigma^2+|\Delta|^2}
+\ln\frac{\bar{M}^2}{\sigma^2}\right)\right]=0
\end{equation}
and
\begin{equation}
|\Delta|\left[1-\frac{4H_S}{\pi}\ln\frac{\bar{M}^2}{\sigma^2+|\Delta|^2}\right]=0.
\end{equation}
It is seen from Eqs.(21) and (22) that the non-zero solutions of
$\sigma$ and $|\Delta|$ will depend on the coupling constants
$G_S$ and $H_S$. Since 2D NG model is an asymptotically free
theory and $G_S$ and $H_S$ can become running, hence it seems that
the non-zero $\sigma$ and $|\Delta|$ will depend on the scale
parameter $\bar{M}$. However, we will prove in Appendix that the
non-zero $\sigma$ and $|\Delta|$ are practically
scale-independent. The same result also arose in the 2D GN model
without the diquark interactions \cite{kn:14} and it simply
reflects the characteristic of the mean field approximation.\\
\indent For discussion of the feature of the extreme value points
of $V(\sigma,|\Delta|)$ we define the determinant
\begin{equation}
K=\left|\begin{array}{cc}
  A & B \\
  B & C \\
\end{array}\right|,
\end{equation}
where
\begin{widetext}
\begin{eqnarray}
 A&=&\frac{\partial^2V(\sigma,|\Delta|)}{\partial \sigma^2}
 =\frac{1}{2G_S}\left(1-\frac{4G_S}{\pi}\ln\frac{\bar{M}^2}{\sigma^2+|\Delta|^2}
 -\frac{2G_S}{\pi}\ln\frac{\bar{M}^2}{\sigma^2}\right)
 +\frac{2}{\pi}\left(1+\frac{2\sigma^2}{\sigma^2+|\Delta|^2}\right) \nonumber\\
 B&=&\frac{\partial^2V(\sigma,|\Delta|)}{\partial\sigma\partial|\Delta|}
 =\frac{\partial^2V(\sigma,|\Delta|)}{\partial|\Delta|\partial\sigma}
 =\frac{4}{\pi}\frac{\sigma|\Delta|}{\sigma^2+|\Delta|^2}
  \nonumber \\
  C&=& \frac{\partial^2V(\sigma,|\Delta|)}{\partial |\Delta|^2}=
  \frac{1}{2H_S}\left(1-\frac{4H_S}{\pi}\ln\frac{\bar{M}^2}{\sigma^2+|\Delta|^2}
  \right)+\frac{4}{\pi}\frac{|\Delta|^2}{\sigma^2+|\Delta|^2}.
\end{eqnarray}
\end{widetext}
Eqs.(21) and (22) have the following four different solutions
which will be discussed successively. \\
1) ($\sigma,|\Delta|$)=(0,0). In this case we can calculate and
get that
$$A=-\infty,\; B=0,\; C=-\infty \;\mathrm{and} \; K=\infty^2>0,$$
hence (0,0) is a maximum point of $V(\sigma,|\Delta|)$.\\
2) ($\sigma,|\Delta|$)=($\sigma_1$,0). By means of the equation
obeyed by non-zero $\sigma_1$
$$1-\frac{6G_S}{\pi}\ln\frac{\bar{M}^2}{\sigma_1^2}=0,$$
it is easy to obtain that
$$A=\frac{6}{\pi},\;\;K=\left(\frac{1}{2H_S}-\frac{1}{3G_S}\right)A.$$
Obviously, ($\sigma_1$,0) is a minimum point of $V(\sigma,
|\Delta|)$ if $G_S/H_S>2/3$ and it will be neither a maximum nor a
minimum point of $V(\sigma,|\Delta|)$ if $G_S/H_S\leq 2/3$.\\
3) ($\sigma,|\Delta|$)=(0,$\Delta_1$). Now by using the equation
satisfied by non-zero $\Delta_1$
$$1-\frac{4H_S}{\pi}\ln\frac{\bar{M}^2}{\Delta_1^2}=0,$$
we get
$$A=\frac{1}{2G_S}-\frac{3}{4H_S}+\frac{2}{\pi}-
\left.\frac{1}{\pi}\ln\frac{\Delta_1^2}{\sigma^2}\right|_{\sigma\rightarrow
0},\;\; K=\frac{4}{\pi}A.$$ If $G_S\neq 0$ we will have that
$A=-\infty$, $K=-\infty$ thus (0,$\Delta_1$) is not an extreme
value point of $V(\sigma, |\Delta|)$. Only if $G_S=0$, it is just
possible that $A>0$ and $K>0$ thus (0,$\Delta_1$) becomes a
minimum point of $V(\sigma,|\Delta|)$. It is indicated that when
$G_S=0$ we will have $\sigma\equiv 0$ by Eq.(9) and
$V(\sigma,|\Delta|)$ will reduce to $V(0,|\Delta|)$ with the
single order parameter $|\Delta|$ coming from the pure diquark
interactions. \\
4) ($\sigma$,$|\Delta|$)=($\sigma_2$,$\Delta_2$), where the
non-zero $\sigma_2$ and $\Delta_2$ are solutions of the equations
\begin{equation}
1-\frac{2G_S}{\pi}\left(2\ln\frac{\bar{M}^2}{\sigma_2^2+\Delta_2^2}+
\ln\frac{\bar{M}^2}{\sigma_2^2}\right)=0,
\end{equation}
\begin{equation}
1-\frac{4H_S}{\pi}\ln\frac{\bar{M}^2}{\sigma_2^2+\Delta_2^2}=0.
\end{equation}
By means of the two equations we may obtain
\begin{eqnarray*}
  A&=&\frac{2}{\pi}\left(1+\frac{2\sigma_2^2}{\sigma_2^2+\Delta_2^2}\right)>0 \\
  K&=&\frac{8}{\pi^2}\frac{\Delta_2^2}{\sigma_2^2+\Delta_2^2}>0.
\end{eqnarray*}
Hence ($\sigma_2$, $\Delta_2$) is certainly a minimum point of
$V(\sigma, |\Delta|)$ as long as the solution exists. For seeking
the condition in which the solution ($\sigma_2$, $\Delta_2$)
emerges, we find out from Eqs. (25) and (26) that
\begin{equation}
\sigma_2^2 =
\bar{M}^2\exp\left[\frac{\pi}{2}\left(\frac{1}{H_S}-\frac{1}{G_S}\right)\right]
\end{equation}
\begin{equation}
\Delta_2^2 =
\bar{M}^2\exp(-\frac{\pi}{4H_S})\left\{1-\exp\left[\frac{3\pi}{4}\left(\frac{1}{H_S}-\frac{2}{3G_S}\right)\right]\right\}
\end{equation}
Obviously, the solution ($\sigma_2$, $\Delta_2$) exists only if
$G_S/H_S<2/3$.\\
\indent The total results above can be summarized as follows:
\indent In the 2D GN model given by the Lagrangian (1), the
locations of the minimum points of the effective potential
$V(\sigma, |\Delta|)$ depend on the ratios of $G_S$ and $H_S$ and
will be at
\begin{equation}
(\sigma,|\Delta|)=\left\{\begin{array}{lc}
  (0, & \Delta_1) \\
  (\sigma_2, & \Delta_2) \\
  (\sigma_1, & 0) \\
\end{array}\right.\;\;\mbox{if}\;\;
\left\{\begin{array}{c}
  G_S/H_S=0 \\
  0<G_S/H_S<2/3 \\
  G_S/H_S>2/3 \\
\end{array}\right.
\end{equation}
These results are very similar to the ones obtained in the 4D
two-flavor NJL model with both quark-antiquark and diquark
interactions \cite{kn:8}, except the only difference that in
present model the pure diquark condensate solution $(0,\Delta_1)$
arises only if $G_S=0$ however in the 4D model it emerges in a
limited region of $G_S/H_S$: $0\leq G_S/H_S<f(H_S)<2/3$. The
critical value 2/3 of $G_S/H_S$ below which the diquark
condensates emerge is also due to the same fact that only the
two-colors (red and green) of quarks participate in the diquark
condensates but all the three colors of quarks get into the
quark-antiquark condensates.
\section{Conclusions\label{concl}}
We have proven by effective potential approach that, similar to
the case of the 4D two-flavor NJL model, in the 2D two-flavor GN
model with diquark interactions, the interplay between the
quark-antiquark and the diquark condensates depends on the ratio
$G_S/H_S$ of the coupling constants in scalar quark-antiquark and
scalar diquark channel. In the ground state of the model, only
pure quark-antiquark condensates could emerge if $G_S/H_S>2/3$ and
the two condensates could coexist if $0<G_S/H_S<2/3$. However, the
pure diquark condensates could be formed only if $G_S/H_S=0$,
which is different from the 4D model where the pure diquark
condensates could arise in a finite region of $G_S/H_S$. In any
way, the critical value 2/3 of $G_S/H_S$ below which the diquark
condensates could emerge is the same for the 2D GN model and the
4D NJL model and it represents the ratio of the color numbers of
the quarks participating in the diquark and the quark-antiquark
condensates. This is probably a common characteristic of this kind
of two-flavor four-fermion models. It is also indicated that if
the model is considered as some simulation of 2D QCD and the
four-fermion interactions are assumed to come from the heavy color
gluon exchange interactions $-g(\bar{q}\gamma^{\mu}\lambda^aq)^2\;
(\mu=0,1; a=1,\cdots,8)$ via the Fierz transformation
\cite{kn:11}, then one will obtain $G_S/H_S=4/3$ for two-flavor
and three-color case.  Hence based on the above results, similar
to the 4D NJL model case, we will have only the pure
quark-antiquark condensates surviving and no diquark condensates
could appear in the ground state of the 2D two-flavor GN model in vacuum.\\
\appendix*\section{}
We will prove that in the mean field approximation, in the
considered 2D GN model the non-zero order parameters $\sigma_2$
and $\Delta_2$ satisfying extreme value conditions of the
effective potential $V(\sigma, |\Delta|)$ are scale-independent.
First rewrite Eqs.(27) and (28) coming from the extreme value
equations (25) and (26) by
\begin{eqnarray}
\sigma_2^2&=&\bar{M}^2\exp\left[\frac{\pi}{2}\left(\frac{1}{H_S}
-\frac{1}{G_S}\right)\right],\nonumber \\
\Delta_2^2&=&\bar{M}^2\exp(-\pi/4H_S)-\sigma_2^2,
\end{eqnarray}
then we will replace the coupling constants $G_S$ and $H_S$ by
respective running form. For this end, define the bare couplings
\begin{equation}
G_S^0=G_SM^{2-D}Z_G,\; H_S^0=H_SM^{2-D}Z_H.
\end{equation}
When viewed as independent variables, $G_S^0$ and $H_S^0$ do not
depend on the scale parameter $M$, i.e. they obey the equations
\begin{equation}
\frac{dG_S^0}{dM}=\frac{dH_S^0}{dM}=0.
\end{equation}
By means of Eqs. (2.17) and (2.18) which define $Z_G$ and $Z_H$
and the standard renormalization group method \cite{kn:10}, we may
derive the equations
\begin{eqnarray}
  M\frac{dG_S}{dM} &=&  \bar{M}\frac{dG_S}{d\bar{M}}=-\frac{12}{\pi}G_S^2,\nonumber \\
  M\frac{dH_S}{dM} &=&  \bar{M}\frac{dH_S}{d\bar{M}}=-\frac{8}{\pi}H_S^2,
\end{eqnarray}
noting that $\bar{M}$ differs from $M$ only by a constant factor.
Eq. (A4) will give the running coupling constants
\begin{eqnarray}
\frac{1}{G_S(\bar{M})}&=&\frac{6}{\pi}\ln\frac{\bar{M}^2}{\bar{M_0}^2}+
 \frac{1}{G_S(\bar{M_0})},\nonumber \\
\frac{1}{H_S(\bar{M})}&=&\frac{4}{\pi}\ln\frac{\bar{M}^2}{\bar{M_0}^2}+
 \frac{1}{H_S(\bar{M_0})},
\end{eqnarray}
where $M_0$ is an arbitrarily fixed momentum scale. Now in Eq.
(A1) we make the replacements
$$
G_S\rightarrow G_S(\bar{M}),\; H_S\rightarrow H_S(\bar{M}),
$$
which will lead to the corresponding substitutions
$$
\sigma_2\rightarrow \sigma_2(\bar{M}),\; \Delta_2\rightarrow
\Delta_2(\bar{M}).
$$
Then by using Eq.(A5), it is easy to verify that
\begin{equation}
\sigma_2(\bar{M}^2)=\sigma_2(\bar{M_0}^2),\; \Delta_2(\bar{M}^2)=
\Delta_2(\bar{M_0}^2),
\end{equation}
i.e. $\sigma_2$ and $\Delta_2$ are scale-independent indeed.

\end{document}